\documentclass[12pt]{iopart}

\usepackage{graphics,color}
\usepackage[dvips]{graphicx}
\usepackage{multirow}
\usepackage{array}
\usepackage{bm}
\usepackage{verbatim}

\begin{document}

\title[Enhancement of microbial motility]{Enhancement of microbial motility due to speed-dependent nutrient absorption}

\author{Mario E. Di Salvo and C. A. Condat}
\address{IFEG-CONICET and FaMAF, Universidad Nacional de C\'ordoba, 5000-C\'ordoba, Argentina}

\date{\today}

\begin{abstract}
Marine microorganisms often reach high swimming speeds, either to take advantage of evanescent nutrient patches or to beat Brownian forces. Since this implies that a sizable part of their energetic budget must be allocated to motion, it is reasonable to assume that some fast-swimming microorganisms may increase their nutrient intake by increasing their speed $v$. We formulate a model to investigate this hypothesis and its consequences, finding the steady state solutions and analyzing their stability. Surprisingly, we find that even modest increases in nutrient absorption may lead to a significant increase of the microbial speed. In fact, evaluations obtained using realistic parameter values for bacteria indicate that the speed increase due to the enhanced nutrient absorption may be quite large.
\end{abstract}

\pacs{87.17.Jj, 87.17.Aa, 05.40.Fb}

\noindent{\it keywords\/}: Bacterial dynamics, energetics, self-propelled organism, microorganism motion.


\section{Introduction}
Many microorganisms inhabit aqueous media. Some of them simply move with the environmental flux, while others are capable of swimming, which allows them to find more easily the nutrients they need. The energy fraction dedicated to motion is generally bigger in the smaller motile microorganisms, which suggests that they have evolved in order to optimize the energetic resources used for motility. The relative costs of locomotion have been recently discussed \cite{Taylor12}. Nutrient availability and consumption rates determine the energy balance of an organism and, therefore, they may determine the future of a species in a given environment. Especially in oligotrophic habitats, such as the ocean, microorganisms are forced to efficiently transform the energy they take up from the environment into motional energy. Although research in bacterial physics has been strongly biased to enteric bacteria, such as \emph{E. coli}, there has also been substantial work on the motion of freshwater and oceanic bacteria. Various chemotactic patterns have been identified \cite{Mitchell02,Mitchell06}, and swimming speeds of several hundreds of $\mu$m/s have been recorded in such bacteria as \emph{Thiovulum majus} \cite{Fenchel94}, \emph{``Candidatus Ovobacter propellens"} \cite{Fenchel04}, and the algae-tracking \emph{Shewanella putrefaciens} and \emph{Pseudoalteromonas haloplanktis} \cite{Barbara03}. The interest in marine bacteria has been recently rekindled by the discovery that \emph{Vibrio alginolyticus} executes a distinctive three-step (forward, reverse, and flick) swimming pattern that helps it to rapidly respond to chemical gradients \cite{Stocker11,Xie11}. A wealth of models has also been developed to describe various aspects of bacterial motion \cite{Condat05,Peruani07,Benza08,Lambert10,Garcia11,Romanczuk12,Taylor12,Hu13,Vogel13,Berdakin13}. But bacteria are not the only flagellum-propelled organisms to exhibit high swimming speeds. The relative swimming speeds, measured in bodies per second, of the hyperthermophilic archaea \emph{M. jannaschii} and \emph{M. villosus} are the highest found in nature \cite{Herzog12}. These archaea have also recently been shown to use a ``relocate and seek" motion strategy \cite{Herzog12}. The field of the fluid mechanics of planktonic microorganisms has been recently reviewed by Guasto, Rusconi, and Stocker \cite{Guasto12}.

Consideration of the effect of motion on resource acquisition and usage is particularly important in the case of small motile microorganisms, which need to swim at high speeds in order to defeat fast noise-induced directional changes \cite{Mitchell91, Mitchell95}. In their classical work, Berg and Purcell \cite{Berg77} considered the absorption of particles by receptors located on the surface of a sphere of radius $a$ moving at a constant speed $V_0$ with respect to the surrounding fluid. They found that the increment in the particle current is a monotonically increasing function of the P\'eclet number $Pe=aV_0/D$, where $D$ is the diffusion coefficient of the particles. They concluded that the motion of a small microorganism would not significantly increase its nutrient uptake rate, if the nutrient consists of high diffusivity particles. For a bacterium whose characteristic size is 1 $\mu$m, moving at a speed of 30 $\mu$m/s and absorbing nutrients whose diffusion coefficient is of the order of $10^{-5}$ cm$^{2}$/s, Berg and Purcell found that the motion-generated nutrient absorption is just about 2.5\% of the total nutrient input. In a comprehensive study of nutrient fluxes in the presence of fluid motion, Karp-Boss and coworkers found that the effect of motion was even smaller \cite{Karp-Boss96}. As a result, it is generally assumed that nutrient transport to small microorganisms such as bacteria is dominated by molecular diffusion and that swimming and feeding currents play a negligible role. However, Logan and Kirchman found an increment in the uptake of $\left[^3\mbox{H}\right]$leucine by marine bacteria due to an advective flow field \cite{Logan91} and it has been suggested that \emph{Thiovulum majus} can significantly increase its food supply by swimming \cite{Schulz01}. There is also a wealth of experimental results that show that advection-dependent uptake (ADU) is advantageous for larger microorganisms such as eukaryotic cells. As early as 1976, Canelli and Fuhs demonstrated that phosphorus uptake by \emph{Thalassiosira fluviatilis} diatoms fixed on filters increased with fluid velocity. Gavis \cite{Gavis76} and Ki\o rboe \cite{Kiorboe93} discussed how fluid motion could increase the nutrient uptake of phytoplankton. The work of Langlois and coworkers, who used simple models of flagellum-propelled spheres to examine the role of advection in microorganism feeding, suggests that the presence of the oscillating flagellum can enhance the effect of advection on nutrient uptake \cite{Langlois09}. The effect of the flagellum was also examined by Short \etal, who used the algae \emph{Volvox carteri} to show that advection of fluid by the coordinated beating of surface-mounted flagella generates a boundary layer of concentration of the diffusing nutrient, playing an important role in the enhancement of nutrient uptake \cite{Short06}. Tam and Hosoi studied the generation of feeding currents in biflagellated phytoplanktons, finding that the breaststroke significantly enhances nutrient uptake \cite{Tam11}, while Michelin and Lauga found that the optimal swimming stroke is essentially independent of the P\'eclet number \cite{Michelin11}.

The influence of advection is stronger if the nutrient diffusion coefficient is smaller. For instance, if this coefficient is of the order of $5\times10^{-8}$ cm/s$^2$, the nutrient uptake rate doubles at a speed of only 15 $\mu$m/s. It is thus important to know whether bacteria may feed on high molecular weight molecules. Although bacteria primarily absorb monomers, Confer and Logan showed that bacteria also absorb macromolecules (after being enzymatically hydrolyzed into subunits) and that this uptake can be increased with fluid shear \cite{Confer91}. Confer and Logan also noted that macromolecular compounds may account for an important fraction of dissolved organic matter in natural waters and wastewaters. In fact, Sugimura and Suzuki found that 80\% of the aminoacids in the North Pacific were in compounds with molecular weights above 1.5 kDa \cite{Sugimura88} and Benner indicates that large macromolecules (sizes ranging from $10^{-9}$ to $10^{-6}$ m) make up about 30\% of the dissolved organic carbon in the the ocean surface \cite{Benner11}. Moreover, it was found that 50 to 60\% of the dissolved organic carbon species in wastewaters have molecular weights above 1 kDa \cite{Grady84}. These high molecular weights imply a relatively low diffusivity ($10^{-6}-10^{-7}$ cm$^2$/s) and, consequently, a higher sensitivity of the nutrient absorption rate to the relative speed between microorganism and water. On the other hand, Logan and Hunt \cite{Logan87} and Logan and Dettmer \cite{Logan90} predicted through a mass transfer analysis that, under certain conditions, fluid motion can increase the assimilation of nutrients by attached microorganisms. Later, Logan and Kirchman found that $\left[^3\mbox{H}\right]$leucine uptake rate by marine bacteria fixed on filters was up to eight times higher within an advective flow field \cite{Logan91}. This effect was only observed at low leucine concentrations, when uptake was likely not saturated. Fluid flow past bacteria did not increase $\left[^3\mbox{H}\right]$glucose uptake, however. These results indicate that the increase in the absorption of leucine is affected by causes more complex than a change in the nutrient molecular weight.

In summary, motion may increase substantially the absorption of nutrients by a microorganism, but this effect is a function of many factors, including the type of substrate and its concentration and the nature of the flow around the microorganism. In this work we study the effect of speed-dependent nutrient uptake on the average speed of a microorganism. To do this, we first generalize a model for the active motion of microorganisms \cite{Schweitzer98} to account for the presence of an ADU rate, and then find its steady state solutions, investigating their stability and other properties. We then obtain numerically the behavior of their time-dependent solutions, which clearly exhibit the presence of two very different time scales, one related to the variation in the stored energy and the other to variations in the speed (this was already noted in the case of constant absorption rate \cite{Condat11}). Finally, we evaluate the possible influence of speed-dependent absorption on the steady-state speed of various microorganisms that are candidates to benefit from it.

\section{Methods: The SET Model}

In 1998, Schweitzer, Ebeling and Tilch (SET) introduced the concept of Brownian motion with energy depot to describe the motion of a microorganism that moves under the combined action of Brownian forces and of its own propulsion system \cite{Schweitzer98,Romanczuk12}. Of course, the term ``depot" is not meant to convey the existence of a specific storage location inside the organism, but it is an idealization introduced to represent the totality of its available biochemical energy stores. This work was later extended by Condat and Sibona \cite{Condat02a,Condat02b,Sibona07} and by Garcia \etal \cite{Garcia11}. SET's basic assumption is that the microorganism can take up energy from the environment at a rate $q(v)$, and store it in a internal energy depot whose instantaneous energy content is $E(t)$. Here $v$ is the instantaneous speed of the microorganism and $t$ the time. The stored energy can be either reconverted into kinetic energy, at a rate $D[v,E(t)]$, or dissipated at a rate $G[E(t)]$. This dissipation rate is assumed to account for all the nonmechanical uses of the available energy. The amount of stored energy is therefore described by the equation
\begin{equation} \label{01}
\frac{\rmd E(t)}{\rmd t}=q(v)-G[E(t)]-D[v,E(t)]\,.
\end{equation}

Since we are interested in the relation between energy absorption and speed, the variables $x$ and $v$ will describe motion along the microorganism trajectory; in the case of run-and-tumble bacteria and other broken-trajectory microorganisms, we will consider only motion in the run phase. To account for the Brownian contribution SET postulated a Langevin equation to describe the motion of the microorganism. However, in the cases of fast-swimming microorganisms, noise plays a negligible role since it generates some rotational diffusion during the run but does not substantially affect the speed. In this work we will neglect the effect of noise, postulating the following equation of motion:
\begin{equation} \label{02}
m\frac{dv}{dt}=-\gamma v+\frac{D[v,E(t)]}{v}\,.
\end{equation}

Equations (\ref{01}) and (\ref{02}) are rather general. In order to make concrete predictions we will make some additional assumptions:
\begin{enumerate}
\item The rates of energy conversion and dissipation are both proportional to the depot energy, $D[v,E(t)]=k(v)E(t)$ and $G[E(t)]=cE(t)$.
\item The transformation function, $k(v)$, has a power-law dependence on the speed, $k(v)=d_{\xi}v^{\xi}$. This generates a rather general model, which may account for the behavior of many microorganisms (SET made the Ansatz that the rate of conversion to kinetic energy is proportional to the instantaneous kinetic energy, so that they wrote $k(v)=d_2v^2$). Measurements of the torque-angular speed relationship in various bacteria indicate that the torque $\tau$ generated by the flagellar motor remains approximately constant up to relatively high angular speeds $\omega$ \cite{Chen00,Sowa03,Li06}. This means that the power $P$ supplied to the flagellum, $P=\omega\tau$, is a linear function of the flagellar angular speed. If we further assume that the translational speed $v$ is proportional to the flagellar rotation rate $\omega$, we could conclude that $\xi\approx1$. The relation between $v$ and $\omega$ has been studied by Magariyama and coworkers, who measured simultaneously the swimming speed and the flagellar rotation rate of the monotrichous bacterium \emph{V. alginolyticus} \cite{Magariyama95} and the swimming speed and the flagellar-bundle rotation rate of the peritrichous \emph{S. typhimurium} \cite{Magariyama01}, finding a roughly linear relation between these two quantities ($v\approx\alpha\omega$). A plausible explanation of why molecular motors have evolved in order to exhibit this relationship is given in \cite{DiSalvo12}. This value is also suggested by the form of the low-speed acceleration experienced by the microorganism, which, if $k(v)=d_{\xi}v^{\xi}$ has the form $a\approx(q_0/mc)d_{\xi}v^{\xi-1}$. If $\xi<1$, the low-speed torque required of a bacterial molecular motor would be unphysically large \cite{DiSalvo12}, while the condition $\xi>1$ would lead to weak accelerations. We must remark, however, that, due to the complexity of the propulsion system, the proportionality of $v$ and $\omega$, reasonable for monotrichous bacteria, becomes questionable for multiflagellate bacteria or other microorganisms. On the other hand, measurements by Garcia \etal suggest that values of $\xi$ larger than 1 could be suitable for \emph{Salmonella typhimurium} \cite{Garcia11}. Therefore, it is advisable to proceed with the analysis for arbitrary nonnegative values of $\xi$, which may also be appropriate for microorganisms other than bacteria.
\item The dependence of the absorption rate on the speed will be assumed to be linear, $q(v)=q_0+Av$. This is not only the simplest choice, but it is also in agreement with the experimental findings of Logan's group on leucine absorption \cite{Logan91}.
\end{enumerate}

With these assumptions, we will discuss the dynamics of the system, calculate the steady-state solutions of the equations of motion, and use standard methods \cite{Britton03} to investigate their stability.

\section{Results}

\subsection{Steady States}

The steady state solutions of equations (\ref{01}) and (\ref{02}) satisfy
\numparts
\begin{eqnarray} \label{03}
q_0+Av-cE-d_{\xi}v^{\xi}E=0\, \\
\frac{d_{\xi}v^{\xi-1}E}{m}-\frac{\gamma v}{m}=0\,.
\end{eqnarray}
\endnumparts
The nontrivial steady-state speeds are thus the real, nonnegative solutions $v_s$ of the equation
\begin{equation} \label{04}
f(v)\equiv-\gamma d_{\xi}v_s^{\xi} + A d_{\xi}v_s^{\xi-1} -\gamma c + q_0 d_{\xi}v_s^{\xi-2} = 0\,.
\end{equation}
If $\xi>1$, $v=0$ is a solution and a bifurcation is present. The steady state solutions for the depot energy are $E_0=q_0/c$, which corresponds to $v=0$, and $E_s=v_s^{2-\xi}/d_{\xi}$ for $v=v_s$. From equation (\ref{04}), we see that the solution $v_s$ depends on the parameters $c$ and $d_{\xi}$ only through the combination $c/d_{\xi}$. Analytical expressions for the nontrivial solutions $v_s$ cannot be obtained for arbitrary values of $\xi$, but they are easy to find in a few important particular cases. The nonnegative analytical solutions of equation (\ref{04}) for the special cases $\xi=0$, 1, and 2 are as follows:

\vspace{1cm}
1. If $\xi=0$, $k(v)=d_0$ (a constant) and there is only one physically acceptable stationary speed,
\begin{equation} \label{05}
v_{s}^{(0)}= \frac{Ad_0}{2\gamma(c+d_0)}+\sqrt{\left[\frac{Ad_0}{2\gamma(c+d_0)}\right]^{2}+\frac{q_0d_0}{\gamma(c+d_0)}}\,.
\end{equation}
\vspace{1cm}

2. If $\xi=1$, $k(v)=d_1v$ and, as above, there again is only one physically acceptable stationary speed,
\begin{equation} \label{06}
v_{s}^{(1)}= \frac{A}{2\gamma}-\frac{c}{2d_1}+\sqrt{\left(\frac{A}{2\gamma}-\frac{c}{2d_1}\right)^{2}+\frac{q_0}{\gamma}}\,.
\end{equation}
Both (\ref{05}) and (\ref{06}) are monotonically increasing functions of $A$ and the coefficient $d_\xi$.
\vspace{1cm}

3. If $\xi=2$, $k(v)=d_2v^2$ and we have three possible stationary speeds: the trivial one $v_s=0$ and
\begin{equation} \label{07}
v_{s\pm}^{(2)}= \frac{A}{2\gamma} \pm \sqrt{\left(\frac{A}{2\gamma}\right)^{2}+\frac{q_0}{\gamma}-\frac{c}{d_2}}\,,
\end{equation}
provided that either $Q>1$ or
\begin{equation} \label{08}
A \geq A_c = 2 \sqrt{\frac{\gamma}{d_2}(\gamma c - d_2 q_0)}\,,
\end{equation}
if $Q<1$.

The solution $v_{s+}^{(2)}$ is always stable and it is a monotonically increasing function of both $Q\equiv q_0d_2/(\gamma c)$ and $A$. Parameter $Q$ relates the product of the energy absorption and active motion parameters ($q_0$ and $d_2$) with the product of the external and internal dissipation parameters ($\gamma$ and $c$). The solution $v_{s-}^{(2)}$ is always unstable and the trivial solution is unstable if $Q\geq1$. The above-mentioned experiments of Garcia \etal were performed in the supercritical regime ($Q>1$) \cite{Garcia11}.

\begin{figure}
\centering
\begin{minipage}[t]{1\textwidth}
\begin{center}
\includegraphics[scale=0.3]{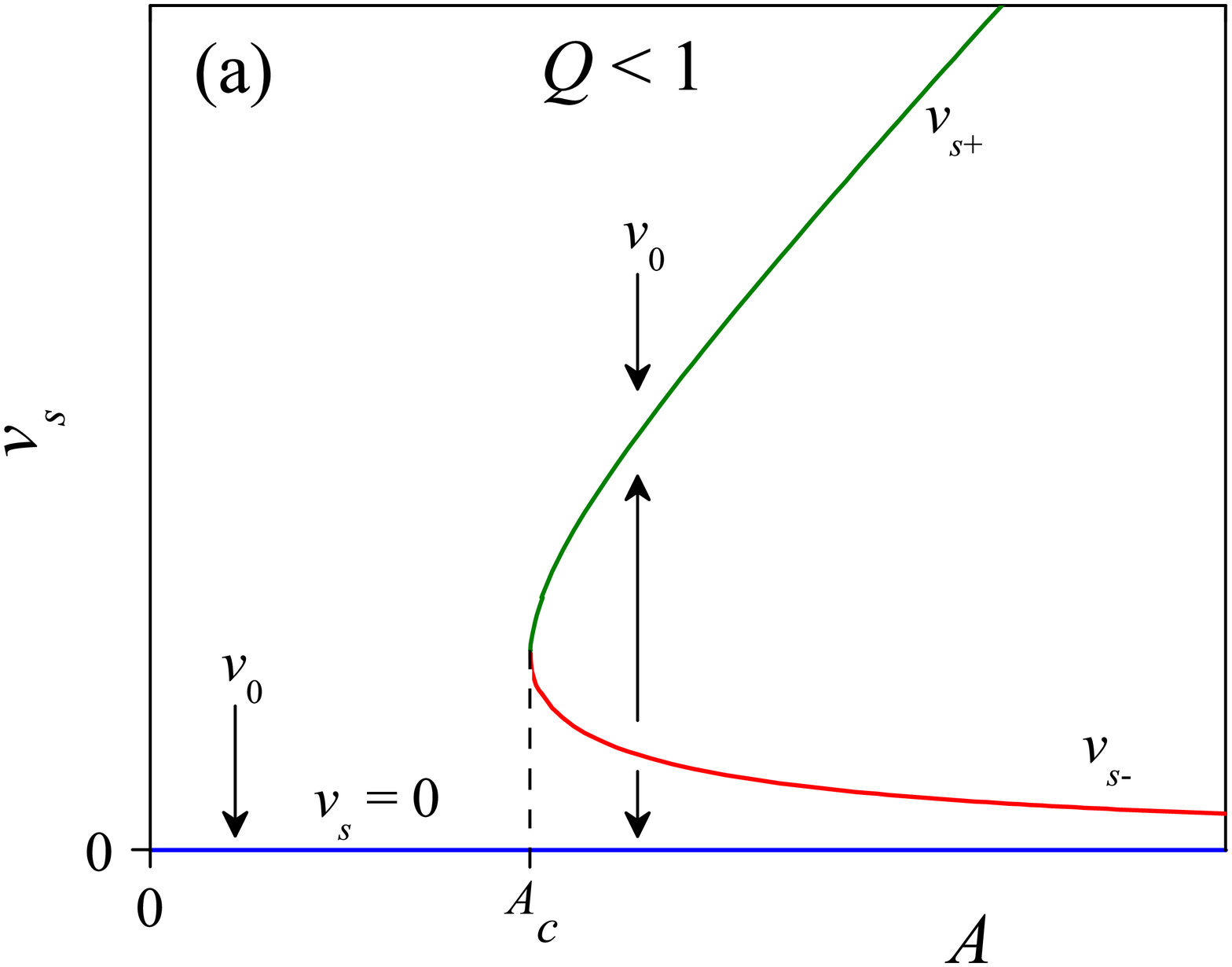}
\end{center}
\end{minipage}
\hfill
\begin{minipage}[t]{1\textwidth}
\begin{center}
\includegraphics[scale=0.3]{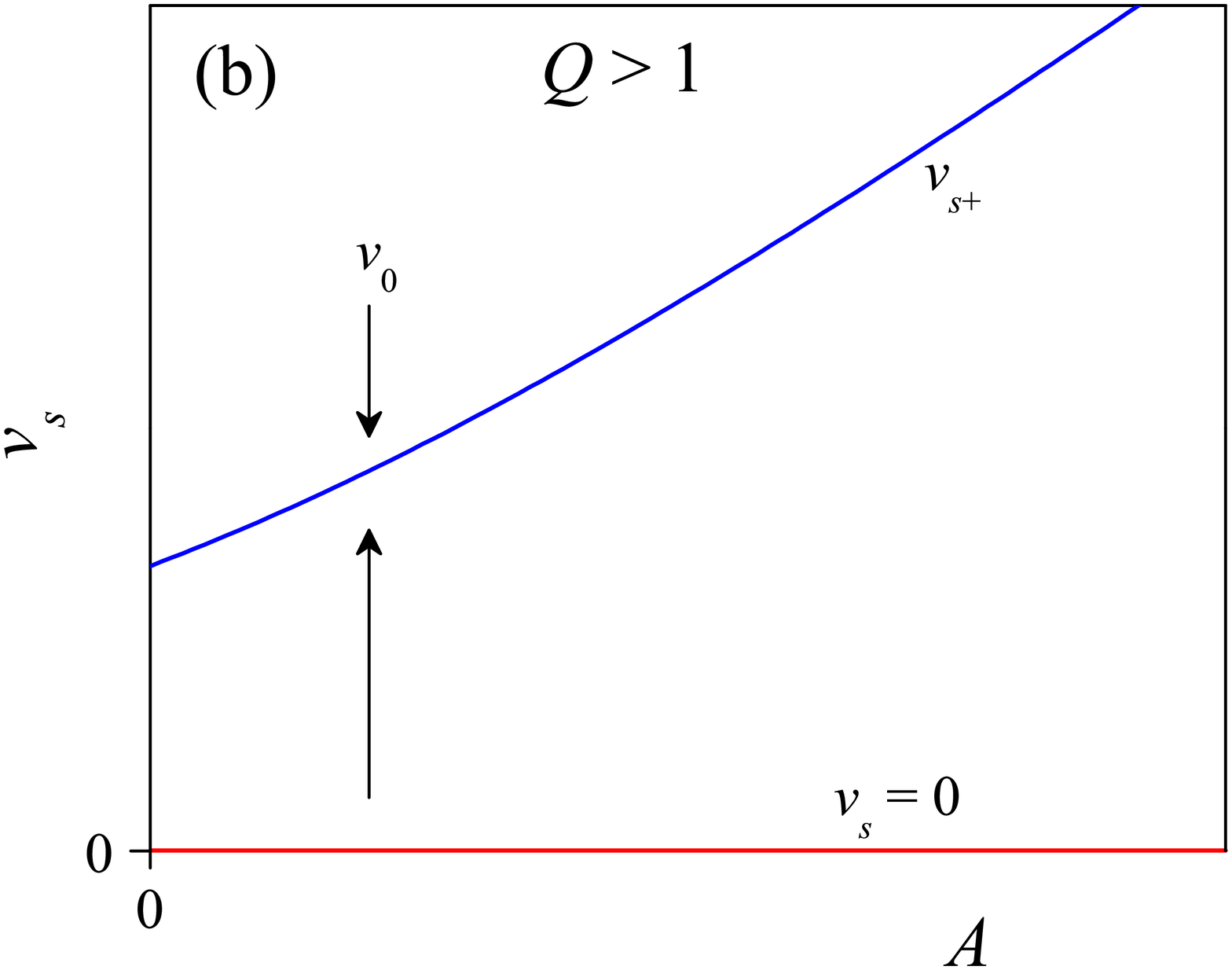}
\end{center}
\end{minipage}
\hfill
\caption{Phase diagram $v_s$ vs ADU strength for $k(v)=d_2v^2$, for the cases (a) $Q<1$ and (b) $Q>1$. The arrows indicate evolution towards a steady state. \label{fig1}}
\end{figure}

The steady-state results can be conveniently presented using phase diagrams in the $A$-$v_s$ plane. Figure \ref{fig1} shows the phase diagrams for the case $\xi=2$ when (a) $Q<1$ and (b) $Q>1$. If the conditions are unfavorable, i.e. high friction or high metabolic consumption ($Q<1$) (the subcritical or high dissipation regime in the language of \cite{Garcia11}), there is a bifurcation at $A=A_c$. For $A<A_c$ only the trivial solution is stable. When $A$ reaches $A_c$, a second stable solution, $v_{s+}^{(2)}$, emerges, which coexists with the still stable trivial solution for $A>A_c$. The unstable solution $v_{s-}^{(2)}$ intervenes between the stable solutions $v_{s+}^{(2)}$ and $v^{(2)}=0$. If $A>A_c$, $v_{s+}^{(2)}$ will be the attractor for any motion whose initial speed $v_0^{(2)}$ is above the separatrix $v_{s-}^{(2)}$, while $v=0$ will be the attractor if the initial microorganism speed is below $v_{s-}^{(2)}$ (all motion ceases in the absence of noise). If the conditions are favorable, i. e. high $q_0$ absorption and small metabolic consumption ($Q\geq1$) the only stable solution is $v_{s+}^{(2)}$, which means that, independently of the value of $v_0^{(2)}$, the speed of the microorganism will always tend to $v_{s+}^{(2)}$.

We first consider that the nutrient uptake is independent of the microorganism speed ($A=0$). For $0<\xi<2$ there is always one nontrivial solution for every value of $q_0$: sustained motion is possible even for very low nutrient uptake ratios. The special case $\xi=2$ was already studied in \cite{Schweitzer98}. These authors found that an absorption rate $q_0>\gamma c/d_2$ ($Q>1$) was required to have a stable nontrivial solution. It is easy to see that this condition can be generalized for arbitrary $\xi\geq2$. We obtain a $\xi$-dependent minimum absorption rate,
\begin{equation} \label{09}
q_c^{(\xi)} = \frac{\gamma \xi}{\xi-2}\left[\frac{c(\xi-2)}{2d_{\xi}}\right]^\frac{2}{\xi}\   ;\   \xi\geq2\,.
\end{equation}
Sustained motion is possible only if nutrient uptake is above this threshold.

Although there is no general closed expression for the threshold $q_c^{(\xi)}$ when $A\neq0$, the size of the minimum amount of nutrient uptake required to keep the microorganism moving can be determined numerically. To do this, two equations must be solved simultaneously:
\numparts
\begin{eqnarray}
f(v*)=0\,  \label{10a} \\
f'(v*)=0\,  \label{10b}
\end{eqnarray}
\endnumparts
with $f(v)$ defined in equation (\ref{04}).

The solutions are presented in figure \ref{fig2}, which was built using realistic values of the parameters (see below). As it could be expected, the threshold shifts to lower values of $q_0$ when $A$ grows. Nutrients obtained through the ADU contribute to keep the microorganism moving. The threshold dependence on the exponent $\xi$ is nontrivial. For low values of $A$, it grows with $\xi$, reaches a maximum and then decreases monotonically to zero. The relevance of this maximum decreases with increasing $A$, until it disappears. Unsurprisingly, $q_c^{(\xi)}$ also decreases monotonically with $A$.

The existence of a maximum in $q_c^{(\xi)}$ for small values of $A$ can be explained by the competition of two different effects: i) Since there is a very fast initial increase in the steady-state speed $v_s$ with $\xi$ (see inset in figure \ref{fig2}), and higher values of $v_s$ demand larger nutrient inputs, $q_c^{(\xi)}$ must be an increasing function of $\xi$ for values of $\xi$ close to two. ii) At higher values of $\xi$, $v_s$ increases slowly with $\xi$, while the fraction of the total absorbed energy devoted to motion increases faster and faster with $\xi$. Increasing $A$ leads to a reduction in the initial increase of $v_s$ with $\xi$ and eventually the maximum disappears.

\begin{figure}
\begin{center}
\includegraphics[scale=0.4]{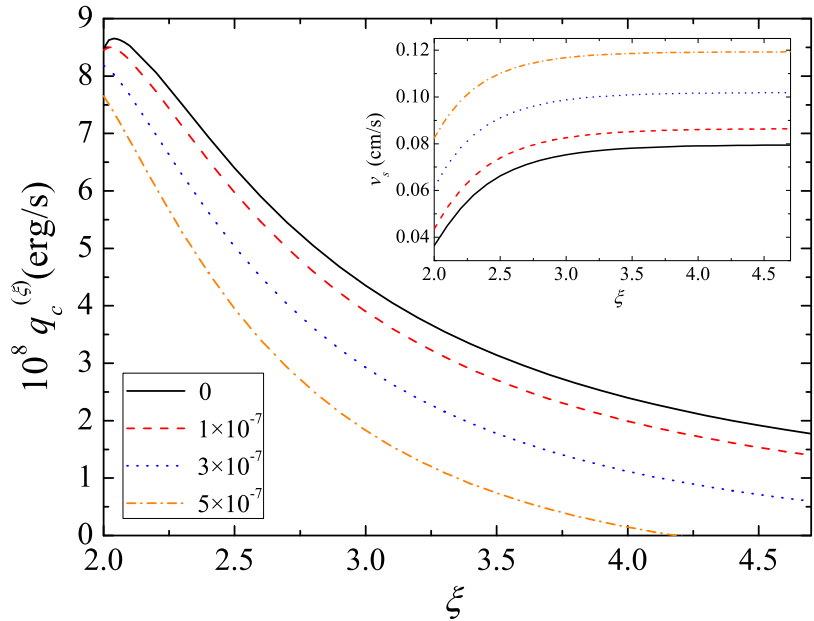}%
\caption{Critical value of the advection-independent absorption rate as a function of the parameter $\xi$ for different values of the ADU strength $A$. If $\xi\geq2$, $q_0$ must always be greater than $q_c^{(\xi)}$ to sustain the motion. Inset: Steady state speed. Here $\gamma=7.54\times10^{-6}$ g/s and $c/d_{\xi}=50\times0.011^{\xi}$ (cm/s)$^{\xi}$  (see below for the parameter choice). \label{fig2}}
\end{center}
\end{figure}

The linear stability of the solutions was investigated using standard methods \cite{Britton03}. Our results are summarized in Table 1. If $0<\xi<2$ there is a single nontrivial stable solution for any values of the nutrient acquisition rate $q_0$; if $\xi \geq 2$, the only stable solution corresponds to a microorganism at rest, unless the advection-independent uptake rate is above the threshold $q_c^{(\xi)}$. These results notwithstanding, we found that in the microbial parameter range, i.e. those parameters found in nature for organisms of very small mass, the nontrivial solution is always a stable focus for $\xi=2$, i.e. the convergence is oscillatory, and unstable for $\xi>2$. Both claims can be easily verified by writing the Jacobian matrix of the set of equations (\ref{01}) and (\ref{02}) and remembering that $\textrm{Tr(J)}<0$ and $\textrm{Det(J)}>0$ implies stability, and that $\textrm{Tr(J)}^2-4\textrm{Det(J)}<0$ implies oscillations.

\begin{figure}
\centering
\begin{minipage}[t]{1\textwidth}
\begin{center}
\includegraphics[scale=0.3]{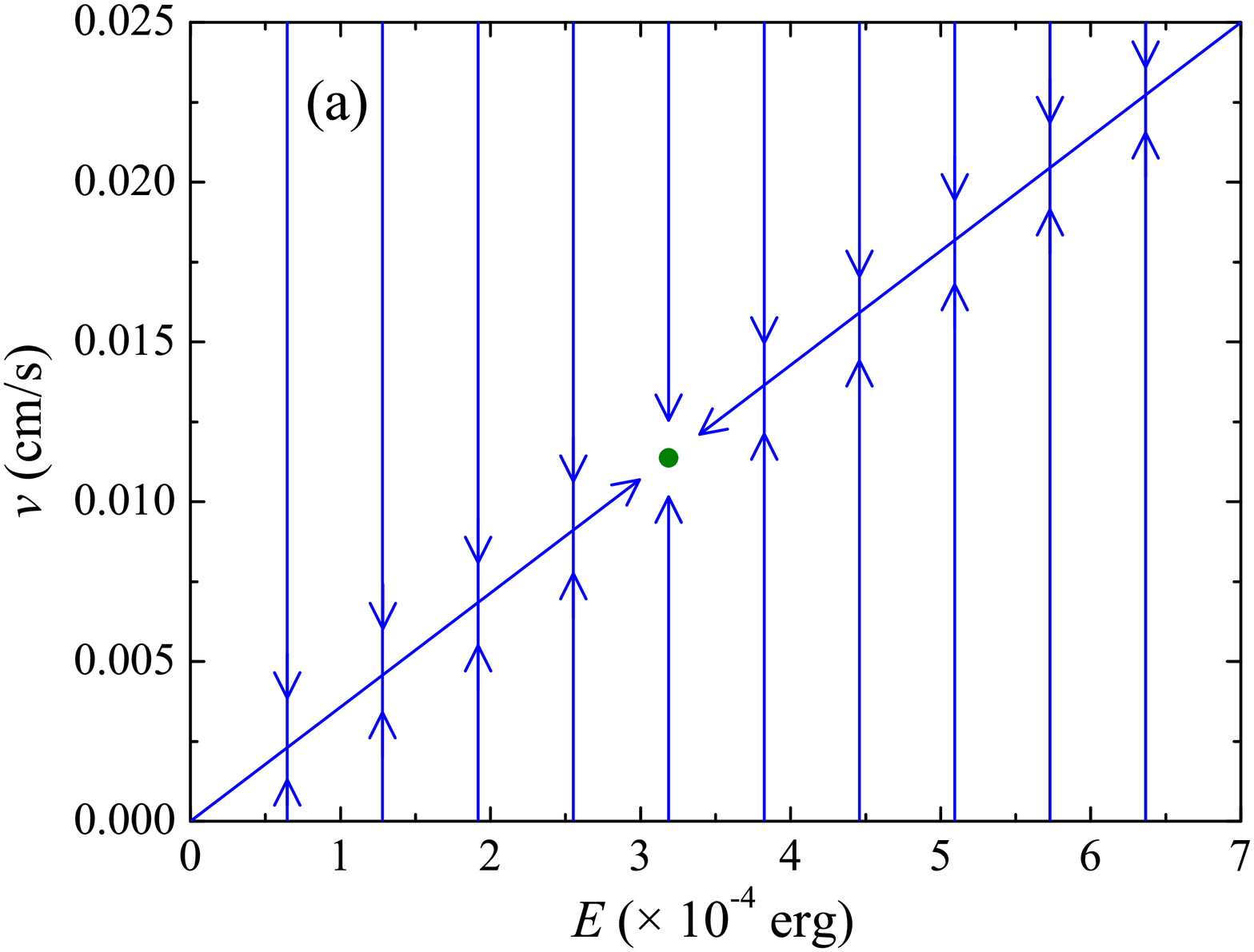}
\end{center}
\end{minipage}
\hfill
\begin{minipage}[t]{1\textwidth}
\begin{center}
\includegraphics[scale=0.3]{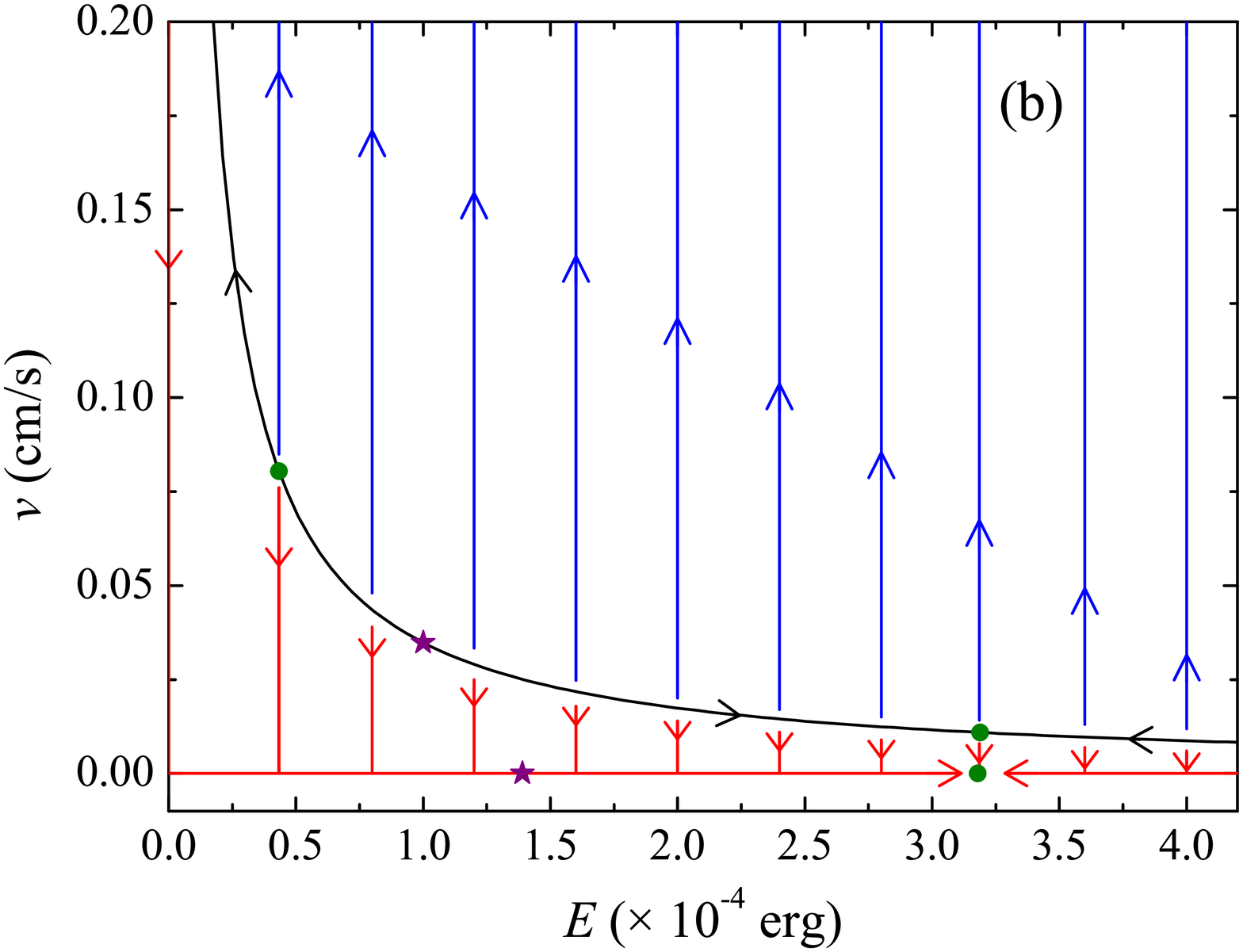}
\end{center}
\end{minipage}
\hfill
\caption{Phase plane representation of the system evolution. (a) $\xi=1$. There is a single stable attractor. Here $q_0=4.77\times 10^{-8}$ erg/s. (b) $\xi=3$. There is a stable attractor corresponding to the microorganism at rest, an unstable fixed point (left dot on the separatrix) and a saddle point (right dot on the separatrix). Purple stars correspond to $q_0=q_c^{(3)}=2.09\times 10^{-8}$ and green dots to $q_0=4.77\times 10^{-8}$ erg/s. In all cases, $\gamma=7.54\times 10^{-6}$ g/s, $c/d_{\xi}=50\times 0.011^{\xi}$ (cm/s)$^{\xi}$ and $A=9.46\times 10^{-8}$ erg/cm.} \label{fig3}
\end{figure}

The system behavior is best shown using the $E$-$v$ phase plane. In figure \ref{fig3}(a) we show the system evolution for $\xi=1$. In this case, there is a single stable node, whose location moves rightward along the oblique line if either $q_0$ or $A$ is increased. The flow lines appear to be straight because, as discussed in \cite{Condat11}, changes in $v$ occur over time scales much shorter than changes in the stored energy: motion along a vertical line usually represents fractions of a second in real time, while a substantial motion along a horizontal (or diagonal) line occurs over times of the order of hours. Panel (b) shows the system evolution for $\xi=3$. There is a single stable attractor, corresponding to the swimmer at rest. The attractor appears initially at the location of the purple star (when $q_0=q_c$) and then moves towards higher energies as either $q_0$ or $A$ is increased. There are two other fixed points, which coincide for $q_0=q_c$ and then separate as $q_0$ or $A$ is increased, both moving along the separatrix in the directions indicated by the arrows. The rightmost fixed point is a saddle point, while the leftmost fixed point is unstable. The separatrix equation is $v_s=\left[\gamma/\left(E\cdot d_{\xi}\right)\right]^{1/(\xi-2)}$.

\begin{table}
\caption{\textbf{Stability of the solutions.} Qualitative features of the stationary solutions for the various ranges of $\xi$.}
\label{table1}
\begin{indented}
\item[]
\begin{tabular}{@{}*{3}{l}}
\br
 &$\boldsymbol{v=0, E_0=q_0/c}$	&$\boldsymbol{v=v_s, E=E_s}$ \\
\mr
\multicolumn{1}{l}{$0\leq\xi\leq1$} &not a solution	&stable \\
\mr
\multicolumn{1}{l}{$1<\xi<2$} &saddle point	&stable \\
\mr
\multicolumn{1}{l}{$\xi=2$} &stable if $Q<1$	&stable focus if $q_0>q_c$ or $A>A_c$ \\
\mr
\multicolumn{1}{l}{$2<\xi$} &stable	&unstable \\
\br
\end{tabular}
\end{indented}
\end{table}

\subsection{Efficiency}

SET defined the mechanical efficiency $\sigma$ of the molecular motor as the ratio of the kinetic energy increase due to the motor to the instantaneous energy uptake \cite{Schweitzer98},
\begin{equation} \label{13}
\sigma=\left\langle \frac{D\left[v,E(t)\right]}{q(v)}\right\rangle\,.
\end{equation}

\begin{figure}
\centering
\begin{minipage}[t]{1\textwidth}
\begin{center}
\includegraphics[scale=0.3]{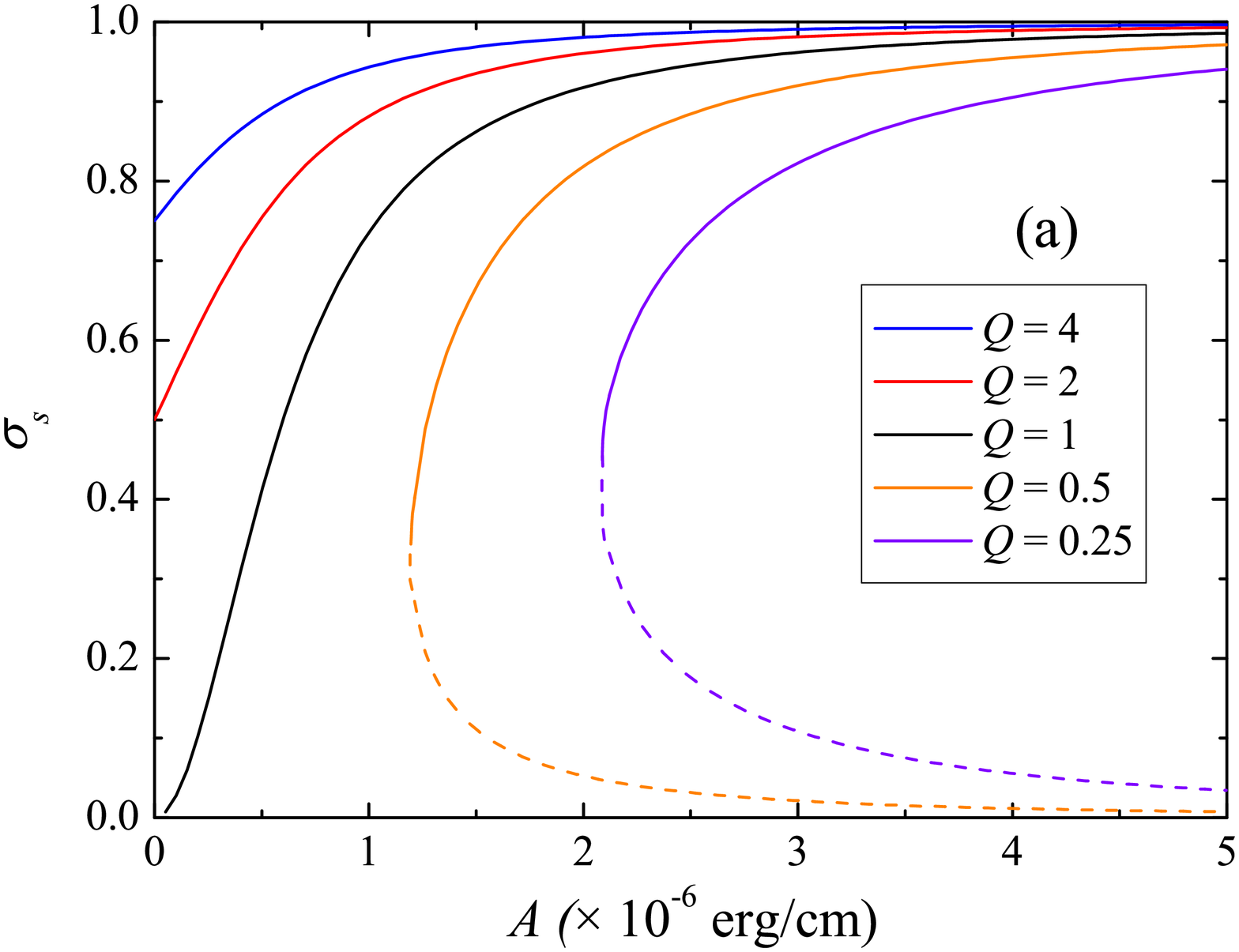}
\end{center}
\end{minipage}
\hfill
\begin{minipage}[t]{1\textwidth}
\begin{center}
\includegraphics[scale=0.3]{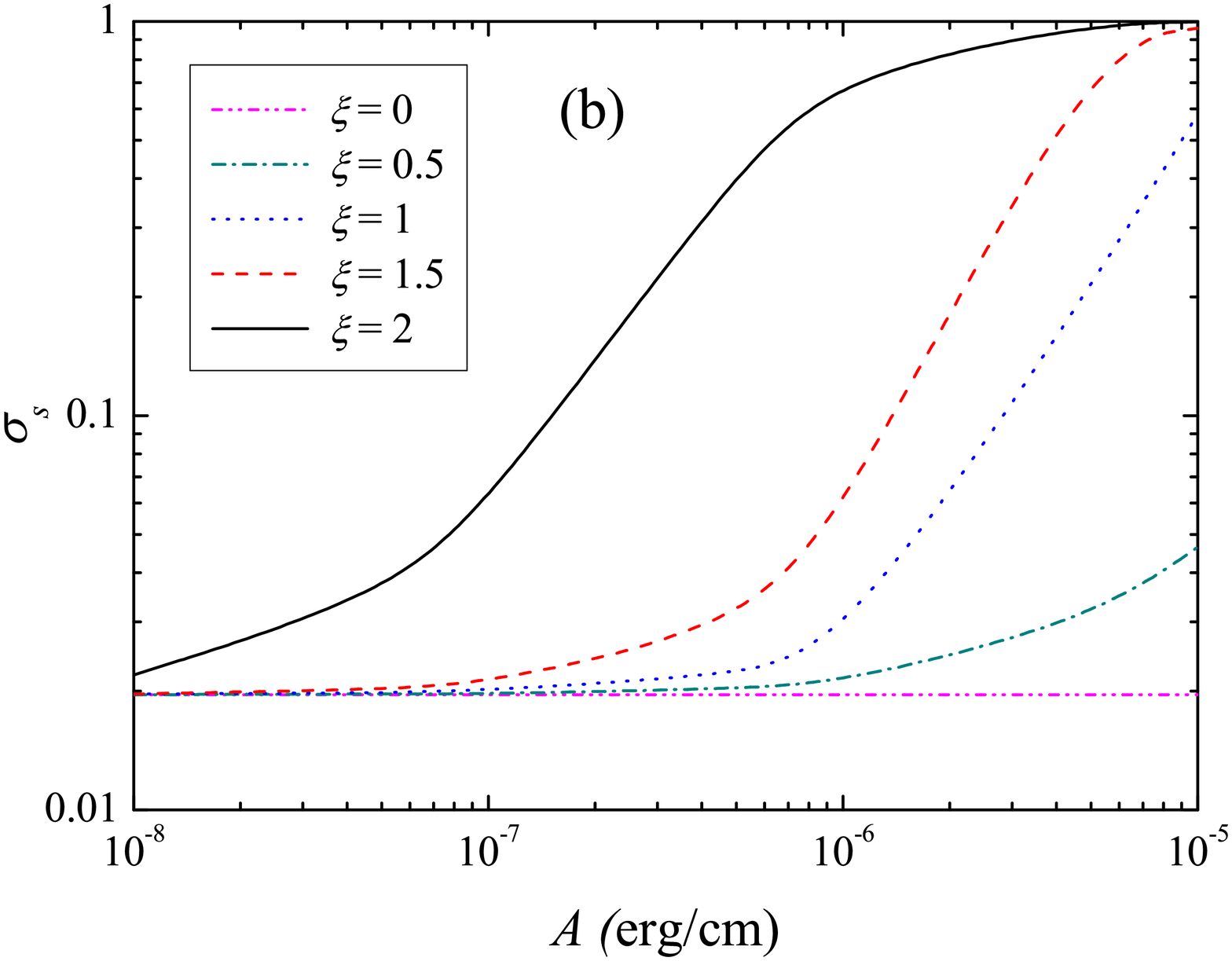}
\end{center}
\end{minipage}
\hfill
\caption{Mechanical efficiency of the molecular motor as a function of $A$ for: (a) $\xi=2$ and the indicated values of $Q$. The dashed lines correspond to the unstable solution $v_{s-}$. (b) $q=4.77\times10^{-8}$ erg/s and different values of the parameter $\xi$. Here $\gamma=7.54\times10^{-6}$ g/s and $c/d_{\xi}=50\times0.011^{\xi}$ (cm/s)$^{\xi}$. \label{fig4}}
\end{figure}

In the absence of noise, and in the model defined by equations (\ref{01}) and (\ref{02}), the steady-state efficiency is thus given by,
\begin{equation} \label{14}
\sigma_s= \frac{d_{\xi}v^{\xi}}{c+d_{\xi}v^{\xi}}\,.
\end{equation}
a form that suggests that $v_s$ may be measured in units of $\left(c/d_{\xi}\right)^{1/\xi}$. The efficiency is always a monotonically increasing function of $v_s^{\xi}$. Equation (\ref{14}) also shows that, as $\xi$ increases, the fraction of the power devoted to motion becomes very small at low speeds, which is consistent with the observation above about the low-speed acceleration. Since the stable solution $v_s^{(\xi)}$ increases with both $q$ and $A$, the efficiency is also a monotonically increasing function of both parameters. Figure \ref{fig4}(a) shows $\sigma$ in the steady state as a function of the parameter $A$ for $\xi=2$ and various values of the parameter $Q$. If $Q<1$ and $A<A_c$ the only possible solution is $v=0$, for which, obviously, $\sigma=0$. At $A=A_c$ a new stable solution, $v_{s+}^{(2)}$, appears, whose efficiency (upper solid line) increases with $A$. The dashed lines correspond to the unstable solution $v_{s-}^{(2)}$. Figure \ref{fig4}(b) exhibits the mechanical efficiency as a function of $A$ for various values of $\xi$. The steady-state efficiency increases with $\xi$, which means that, in the steady-state, a higher value of $\xi$ is more convenient. The reason for this is that, at high speeds, a higher value of the exponent implies that a larger energy fraction can be transformed into motion.

\subsection{The second depot approximation}

In the development of the model, we have assumed that the additional energy resources acquired from advection are distributed between motility and other metabolic functions in exactly the same way as those resources acquired through normal nutrient diffusion. However, if the bacterium goal is to explore space more efficiently, we can assume that a higher proportion of the energy acquired through absorption is devoted to motion. It is not known what this proportion is, but the extreme case would occur when all of the energy acquired through advection is transformed into kinetic energy. It is therefore convenient to consider the two extreme cases: a) the extra power is distributed as in the $A=0$ case, as discussed above; b) all additional power is devoted to motion. Of course, in the latter case we must still allow for the intrinsic inefficiency of the propulsion system.

To study case b) we will assume that all the additional advection-related energy goes to drive the motors. We model this situation by considering that the bacterium has two different energy depots, one that is filled by nutrient diffusion ($E_1$) and another that is filled by the nutrient influx enhancement due to the relative motion between cell and medium ($E_2$). This is a useful artifice but, of course, we are not assuming that the bacterium actually has two ``physical compartments". Under these assumptions, the set of equations (\ref{01}) and (\ref{02}) is replaced by,
\numparts
\begin{eqnarray} \label{15}
\frac{dE(t)}{dt}=q_0-c_1E_1(t)-d_{\xi}v^{\xi}E_1(t)\, \\
\frac{dE(t)}{dt}=Av-c_2E_2(t)-d_{\xi}v^{\xi}E_2(t)\, \\
m\frac{dv}{dt}=-\gamma v+d_{\xi}v^{\xi-1}E(t)\,,
\end{eqnarray}
\endnumparts
where the total energy $E$ is the sum of the energy contents of the two depots ($E=E_1+E_2$). Stability of the steady states can be easily investigated. Considering $q_0\neq0$, we find that there is only one stable solution if $\xi<2$; if $\xi=2$ a bifurcation occurs, and for $\xi>2$ there are two stable solutions, one being $v=0$ and the other, $v_s$, satisfying the equation,
%
%
\begin{eqnarray} \label{18}
\fl f(v) \equiv \gamma d_{\xi}^2v_s^{2\xi} - d_{\xi}^2(Av_s+q_0)v_s^{2\xi-2} + \gamma d_{\xi}(c_1+c_2)v_s^{\xi} \nonumber\\
- d_{\xi}(Av_sc_1+q_0c_2)v_s^{\xi-2} + \gamma c_1c_2 = 0\,.
\end{eqnarray}
As we have shown for the single-depot case, there are two solutions if $q_0$ is larger than a threshold value $q_c$, which is the solution of the set of equations (\ref{10a}) and (\ref{10b}) with $f(v)$ now defined by equation (\ref{18}). The cases for the stability of the solutions are exactly those described before (see Table 1).

\subsection{Microbial dynamics}

To have a realistic perspective we estimate the values of the model parameters for a marine bacterium. Figure 1 in \cite{Logan91} shows a linear growth of the leucine absorption rate with fluid speed. From the slope of the fitting line for speeds between 0 and 230 $\mu$m/s (2.5 (pmol$\cdot$day)/(l$\cdot$min$\cdot$m)) we obtain $A=9.46\times10^{-8}$ erg/cm, assuming that a leucine molecule generates approximately 45 ATP in a prokaryotic cell. We estimate $q_0$ using the theoretical expression for the diffusion flux to the cell surface \cite{Berg77}, $q_0=4\pi Dr(C_{\infty}-C_0)$, where $D$ is the diffusion coefficient of the nutrient, $r$ is the bacterium radius, and $C_{\infty}$ and $C_0$ are the substrate concentrations in bulk water and at the cell surface. According to \cite{Logan91}, these parameters are $r=0.4$ $\mu$m, $D=7\times10^{-6}$ cm$^2$/s, and $C_{\infty}=1$ nM (these are typical values in the ocean) and assuming that the concentration at the cell surface is near zero we get $q_0=4.77\times10^{-8}$ erg/s. On the other hand, taking for the fluid viscosity $\eta=0.01$ g/(cm$\cdot$s), we find $\gamma=6\pi\eta r=7.54\times10^{-6}$ g/s.

Since $cE$ is the power consumed non mechanically and $d_{\xi}v^{\xi}E$ is the net power effectively transformed into motion, we can estimate the fraction $c/d_{\xi}$ by assuming that 20\% of the energy budget is devoted to motion with an efficiency of 0.1 ($d_{\xi}v^{\xi}E/cE=0.02$). It has been estimated elsewhere that the net fraction effectively transformed into kinetic energy is of about 1\% \cite{Mitchell91}, but, since we are only interested in the fastest bacteria, we (conservatively) assume that this fraction is 2\%. It must be noted that this fraction can vary depending on the energetic demands of the bacterium.

\begin{figure}
\centering
\begin{minipage}[t]{1\textwidth}
\begin{center}
\includegraphics[scale=0.3]{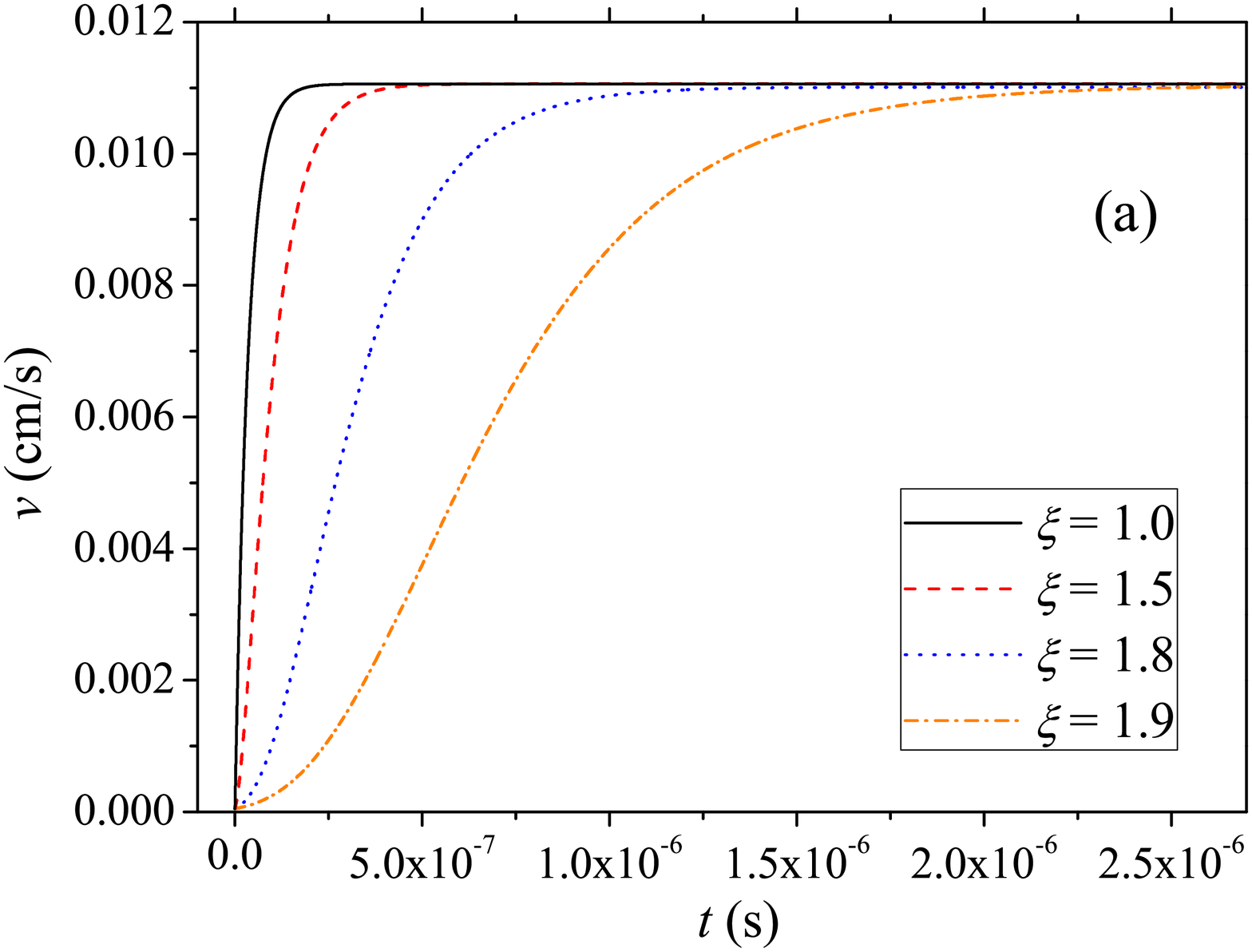}
\end{center}
\end{minipage}
\hfill
\begin{minipage}[t]{1\textwidth}
\begin{center}
\includegraphics[scale=0.3]{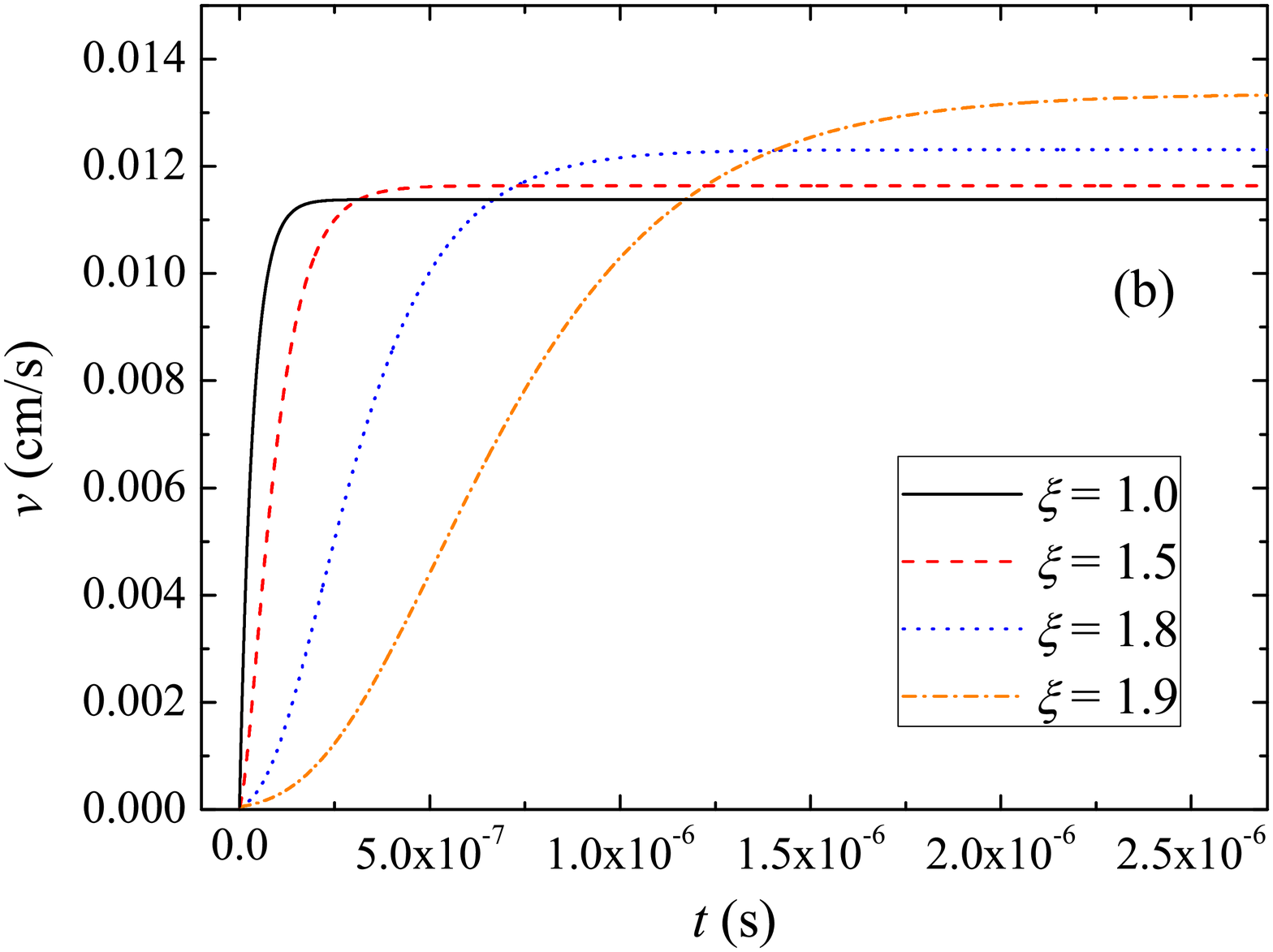}
\end{center}
\end{minipage}
\hfill
\begin{minipage}[t]{1\textwidth}
\begin{center}
\includegraphics[scale=0.3]{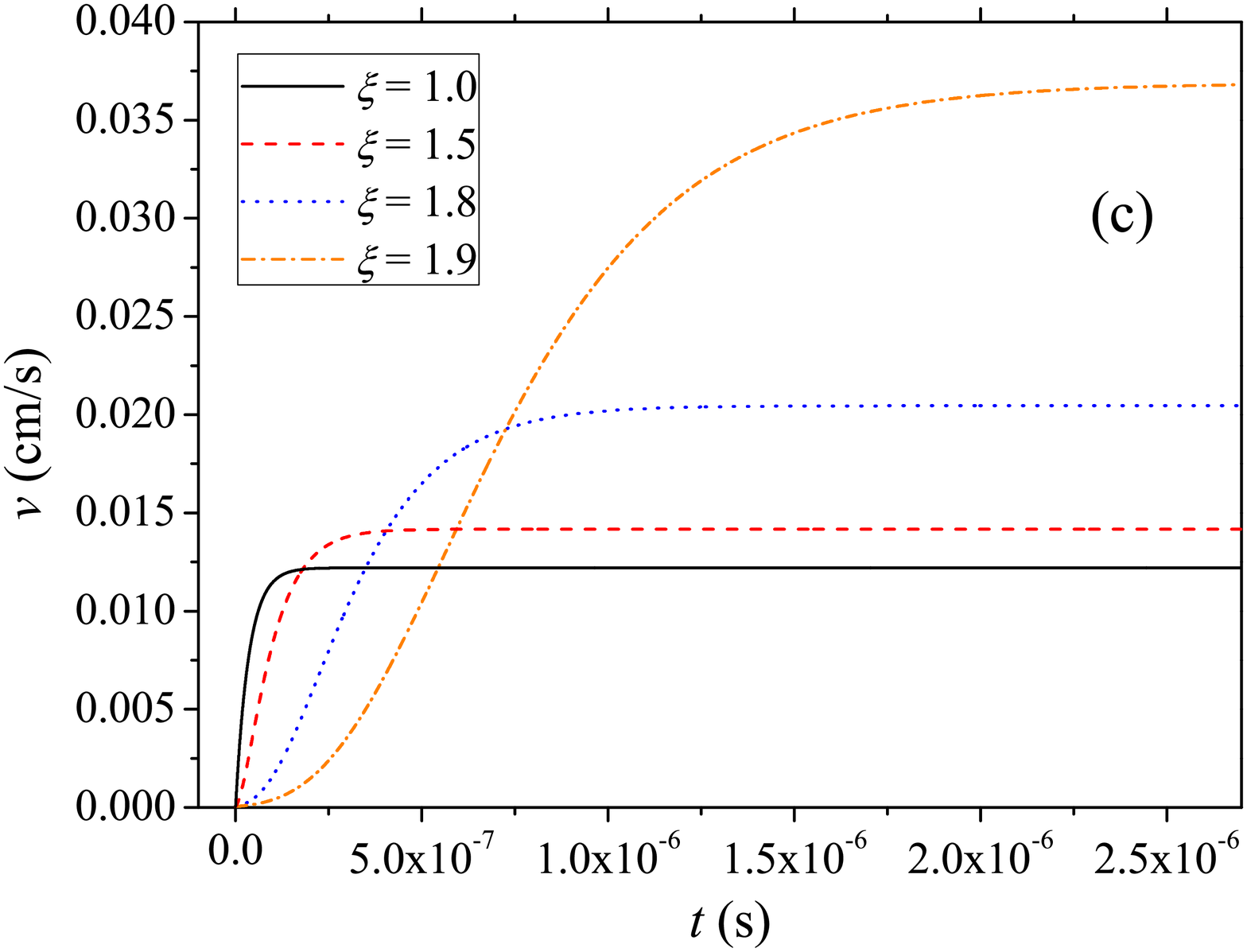}
\end{center}
\end{minipage}
\hfill
\caption{Microbe speed as a function of time for the indicated values of the exponent $\xi$. a) No ADU. b) ADU as in \cite{Logan91}, single depot. c) ADU as in \cite{Logan91}, two depots. Parameter values are chosen as indicated in the text. \label{fig5}}
\end{figure}

To investigate the consequences of changing the exponent $\xi$, we fix the value of $v$ in the absence of ADU to be equal to 110 $\mu$m/s, a characteristic value of the speed for many aquatic microorganisms. For the two-depot case we take $c_1=c$ and $c_2=c/5$; for $c_2$ we only need to consider the inefficiency of the motor because we are assuming that all the energy of the second depot is devoted to motion. Although the parameter values we use are only estimative, they should suffice for the purposes of investigating the effects of varying $A$ and $\xi$. With this choice of parameters, we solved equations (\ref{01}) and (\ref{02}) and plotted the speed of the bacterium as a function of time for both the single- and two-depot cases. The results for $A=0$ are shown in figure \ref{fig5}a, where we see that the fastest acceleration occurs for the smallest values of $\xi$. If $A\neq0$, the initial acceleration is still stronger for $\xi$ close to unity, but now the steady-state swimming speed increases monotonically with $\xi$ (see figure \ref{fig5}b). In the two-depot problem (figure \ref{fig5}c), we see a huge increase in the steady-state speed, especially for values of $\xi$ close to two. In both cases, the microorganism profit more of the ADU for large values of $\xi$.

The P\'eclet number, measures the relative importance of advection versus diffusion. If $Pe<1$, diffusion outcompetes transport by advection from the flowing medium, whereas if $Pe>1$, advection dominates. For very small microorganisms, such as bacteria, $Pe$ is usually of the order of $10^{-2}-10^{-1}$. For these cases, advection is therefore generally assumed to be negligible. In Table 2 we give $Pe$ for three different types of bacteria. It must be noted that the small $Pe$ assumption is satisfied by \emph{E. coli} but fails for fast bacteria (such as \emph{T. majus} or \emph{O. propellens}). What happens for bacteria with $\textrm{Pe} \sim 10^{-1} - 1$? The enhancement of the total nutrient flux to the cell, relative to the purely diffusive nutrient flux to the cell is measured by the Sherwood number, $Sh$. A relation between $Pe$ and $Sh$ has been established by empirical formulae; for instance, Clift et al. proposed $Sh=\frac{1}{2}[1+(1+2Pe)^{1/3}]$ to fit data for a sinking sphere \cite{Clift78}. According to these formulae, a marine bacterium whose radius is 0.5 $\mu$m, moving at a speed of 200 $\mu$m/s and absorbing glucose, has a P\'eclet number of 0.33 and a Sherwood number of 1.09. Other formulae were obtained by analytical methods \cite{Karp-Boss96,Hondzo07}. However, since these microorganisms are not sinking spheres but self-propelled swimmers pushed through the water by one or more beating flagella, they generate feeding currents and the streamlines come closer to the surface of the cell. Considering this effect, $Sh$ is approximately 1.2 for the marine bacterium \cite{Langlois09}, 20\% more than taking into account only diffusion. Table 2 contains the P\'eclet and Sherwood numbers for three bacterial types for different nutrient sizes; $Sh$ was calculated using Clift's formula. The radii and speeds are expressed in $\mu$m and $\mu$m/s, respectively. The diffusion coefficients of the nutrients are: Dissolved organic matter (DOM) $D=6\times10^{-7}$ cm$^2$/s, glucose $D=3\times10^{-6}$ cm$^2$/s and leucine $D=7\times10^{-6}$ cm$^2$/s. The Sherwood number increases with the relative speed of the microorganism with respect to the fluid and with the size of the absorbed nutrient particle.

In the previous example we have taken $q_0\sim5\times10^{-8}$ erg/s, $A\sim10^{-7}$ erg/cm, and $v\sim0.01$ cm/s. In this case, the contribution of advection is of only 2\% of the total absorption, 10 times smaller than the potential ADU. It is important to note that even this small contribution has a non-negligible effect on the final speed of the microorganism. In the Discussion we present a plausible explanation of this discrepancy.Of course, the greater the ADU contribution, the stronger the motility enhancement. We also remark that a marine bacterium can increase its uptake by 30\% if the nutrient absorbed is high-molecular-weight DOM).

\begin{table}
\caption{P\'eclet and Sherwood numbers for some bacteria.}
\label{table2}
\lineup
\begin{indented}
\item[]
\begin{tabular}{@{}*{9}{l}}
\br
\multicolumn{1}{l}{Bacterium} &Radius &Speed &\centre{2}{DOM} &\centre{2}{Glucose} &\centre{2}{Leucine} \\
\multicolumn{1}{l}{} & & &$Pe$ &$Sh$ &$Pe$ &$Sh$ &$Pe$ &$Sh$ \\
\mr
\multicolumn{1}{l}{Marine} &0.5 &200 &\01.67 &1.32 &0.33 &1.09 &0.14 &1.04\\
\mr
\multicolumn{1}{l}{\emph{E. coli}} &1 &\020 &\00.33 &1.09 &0.07 &1.02 &0.03 &1.01\\
\mr
\multicolumn{1}{l}{\emph{O. propellens}} &2 &600 & 20.00&2.22 &4.00 &1.54 &1.71 &1.32\\
\br
\end{tabular}
\end{indented}
\end{table}

\subsection{Motility enhancement}

In order to assess the influence of advection-dependent absorption on the stationary speed of a microorganism, we define the function motility enhancement, $\Psi^{(\xi)}(A)$, as the ratio of the stationary speed when nutrient absorption is speed-dependent to the stationary speed when nutrient absorption is constant, for every value of the parameter $\xi$,
\begin{equation} \label{13}
\Psi^{(\xi)}(A)=\frac{v_s^{(\xi)}(A\neq0)}{v_s^{(\xi)}(A=0)}
\end{equation}

\begin{figure}
\begin{center}
\includegraphics[scale=0.4]{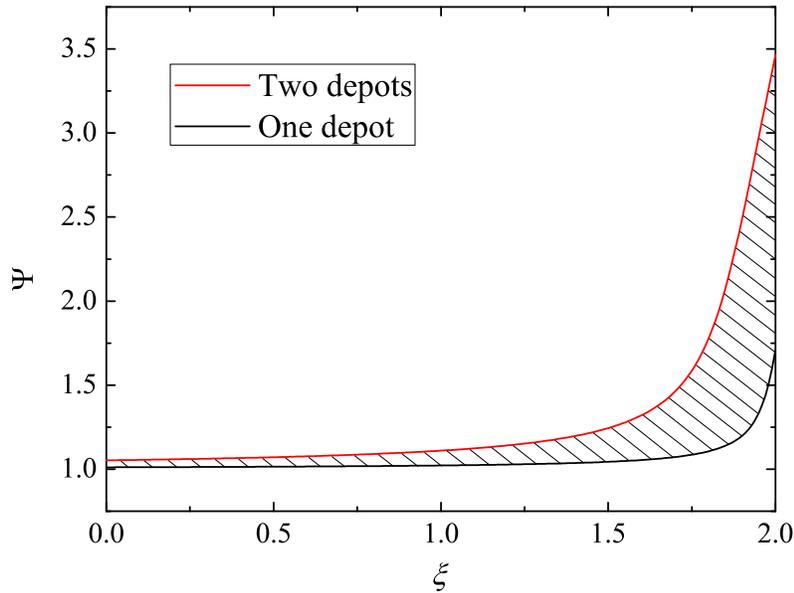}%
\caption{Motility enhancement due to speed-dependent nutrient uptake as a function of the parameter $\xi$ for the two limiting cases: when all the ADU energy is distributed uniformly between motility and metabolism (one depot - lower curve) and when all the ADU energy is intended for the motors (two depots - upper curve). The true motility enhancement should be in the hatched region between these two curves. Parameter values are chosen as indicated in the text. \label{fig6}}
\end{center}
\end{figure}

This function yields the increment in motility due to the putative speed-dependent nutrient uptake. For a realistic analysis we use the parameters found above and calculate numerically the solutions of (\ref{04}) (for one depot) and (\ref{18}) (for two depots) for $A=0$ and $A\neq0$, plotting the fraction $\Psi$ as a function of the parameter $\xi$ (figure \ref{fig6}). The plot was prepared with a value of $A$ measured for the absorption of leucine by a small bacterium \cite{Logan91}. The effect for larger microorganisms or for dissolved organic matter of higher molecular weight would be more intense. The fact that $\Psi$ is greater than one means that it is always beneficial for the microorganism to have an absorption rate proportional to the speed. The contribution of the speed-dependent nutrient uptake to the final speed becomes very important as $\xi$ nears 2.

\section{Discussion}

In this paper we have introduced a generalization of the SET model to take into account the advective contribution to the microbial nutrient uptake. We have incorporated this effect by adding to the absorption rate a term that is proportional to the microorganism speed, in agreement with the figure 1 of \cite{Logan91}. Since we have used realistic values of the parameters involved, this procedure allows us to investigate the problem quantitatively as well as qualitatively. We have considered two limiting cases: A) Additional energy proportionally distributed between motion and other uses. B) Additional energy totally devoted to motion. It is reasonable to assume that an intermediate situation is likely to occur in nature, and that motility will be enhanced by a proportion indicated by some point in the hatched region of figure \ref{fig6}.

Although the effect of advection on small microorganisms such as bacteria has been usually ignored, our results suggest that, even a small advection-related contribution to the uptake could have a significant effect on the motility of a microorganism. In particular, both the final speed of the microorganism and the steady-state efficiency of its molecular motors may be considerably increased.

As an example, we can evaluate the contribution of advection to the final speed for \emph{E. coli} and \emph{O. propellens}. According to equation (\ref{13}) this contribution depends on the parameter $\xi$. Concretely, we will consider the exponents $\xi=1$ and 1.5 and the one-depot case. If the absorbed nutrient is glucose, we find that $Sh$ is 1.02 and 1.54 for \emph{E. coli} and \emph{O. propellens}, respectively, as it can be seen in Table 2. For \emph{E. coli}, the increment in the final speed is a negligible 2\% for $\xi=1$ and 12\% for $\xi=1.5$; for \emph{O. propellens}, the predicted increment in the final speed would be 25\% and 240\%, respectively.

Some points deserve further discussion:

\emph{The experiment of Logan and Kirchman} \cite{Logan91}. As far as we know, this is the only direct experimental measurement of the effect of flow on bacterial nutrient uptake. In most of our examples, we used the value of $A$ extracted from this experiment. Logan and Kirchman obtained a linear relation between nutrient absorption and speed, when nutrient concentration is not saturated. The experiment was performed using attached cells; since these cells were not actually swimming, it is possible that some nutrient receptors were blocked, that the fluid stream lines were different from those resulting from true swimmers or that the cell energetic demands were reduced because the microorganisms were not swimming. What is remarkable is that, even with this modest value of the parameter $A$, we could observe that the effect of ADU is non-negligible even for very small microorganisms, such as bacteria.

\emph{Small speed regime: the choice of an optimal $\xi$}. From equation (\ref{02}) we concluded that, if $D\left[v,E(t)\right]=d_{\xi}vE(t)$, the leading contribution to the low-speed acceleration experienced by the microorganism has the form $a\approx(q_0/mc)d_{\xi}v^{\xi-1}$. Since the acceleration must be finite, but not too small, we can thus argue that, at low speeds, $\xi$ must be close to unity: In the absence of noise, $\xi=1$ leads to constant acceleration, while $\xi<1$ and $\xi>1$ would require, respectively, enormous torques and very long speed-up times. This can also be seen from figure \ref{fig5}.

\emph{High- speed regime: the virtuous circle.} At high speeds, stronger accelerations and a bigger increment in the final speed of the microorganism would result from higher values of $\xi$. A higher speed leads to higher absorption that leads to higher speeds, and so on. The stationary speed increases with $A$, so it is possible that the microorganism could activate additional uptake channels in order to enhance the ADU and, in this way, be able to "jump" to a higher speed steady state. Our calculations indicate that fast motion would be favored by a shift to a higher $\xi$ state. We ignore if nature avails itself of this possibility, but it would be very interesting to have experimental information that tests it.

\emph{Prediction generality}. Even though we chose bacteria for the numerical examples, the analysis need not be restricted though we particularized the analysis of the model for bacteria it need not be restricted to this type of organisms; the model is very general regarding the size and shape of the cells and it can be applied to study the motion of other types of self-propelled microorganisms. Our model suggests that the speeds of the microorganisms can be considerably enhanced by the presence of speed-dependent absorption; however these outcomes should be contrasted with experimental results. We thus propose further direct measurements of the ADU, perhaps with experiments similar to that of Logan and Kirchman; we suggest using higher molecular weight nutrients in different concentrations and other types of microorganisms.

\ack{We are thankful to Prof. James G. Mitchell for very helpful correspondence at the start of this work and to Dr. G. Sibona for illuminating discussions. This work was supported by SECyT-UNC (Project 30720110100436) and CONICET (PIP 112-200801-00772) (Argentina).}


\end{document}